\documentclass[preprint,showkeys, preprintnumbers,amssymb]{revtex4}

\usepackage{graphicx}
\graphicspath{{figs/}}
\begin{document}
\title{Striped morphologies induced by magnetic impurities in
d-wave superconductors}
\author{Xian-Jun Zuo}
\thanks{Electronic mail: xjzuo@yahoo.com.cn}
	\affiliation{Department of Physics, Zhejiang University of Technology,
 Hangzhou 310023, People¡¯s Republic of China }

\date{\today}

\begin{abstract}
We study striped morphologies induced by magnetic impurities
in d-wave superconductors (DSCs) near optimal doping by
self-consistently solving the Bogoliubov-de Gennes equations based
on the $t-t^{\prime }-U-V$ model. For the single impurity case, it
is found that the stable ground state is a modulated checkerboard pattern.
For the two-impurity case, the stripe-like structures in order parameters are induced due to the impurity-pinning effect.
The modulations of DSC and charge orders share the same period of four lattice constants (4$a$),
which is half the period of modulations in the coexisting spin order.
Interestingly, when three or more impurities are inserted, the impurities
could induce more complex striped morphologies due to quantum interference.
Further experiments of magnetic impurity substitution in DSCs are
expected to check these results.
\end{abstract}

%\pacs{74.20.-z, 74.62.Dh, 74.25.Jb}

\maketitle

\section{introduction}
Recent studies of spatial inhomogeneous phases with stripe or checkerboard
modulations in copper oxide-based compounds has received much attention.
Neutron scattering (NS)experiments reveal the existence of
incommensurate magnetic peaks in cuprates,
which has led to discussions of a stripe phase~\cite{lake1,tranq2,yamada3,mook4,fujita5,hoff6,howa7,hana9,chen6,franz7,chenhy8}.
Meanwhile, scanning tunneling microscopy (STM) experiments observed
checkerboard-like charge-density wave (CDW) modulations~\cite{Mc9,vers10,Kohs11}, which break the square symmetry
of the underlying lattice. It was
proposed that the effect can be understood in terms of quasiparticle
interference (QPI) due to scattering on impurities and other
inhomogeneities~\cite{zhujx12,wangqh13,zhangd14,Podolsky15} within the so-called 'octet' model, or in terms of static or fluctuating
stripes~\cite{kivelson16}. Kohsaka et al. suggest that both dispersive and non-dispersive modulated patterns originate
from different regions in momentum and energy space~\cite{Kohs11}. The well-defined states in the nodal region are responsible for the low energy QPI
structure, whereas the ill-defined quasiparticle states in the antinodal regions are responsible for the non-dispersive charge order above some energy
scale. On the other hand, zero-field experiments on the well ordered YBCO
samples show no evidence of static broken translational symmetry in
nuclear magnetic resonance or NS experiments, which leads to the
proposal that the checkerboard patterns observed by STM is caused by
the local disorders~\cite{yang17}. In brief, the physics that determines the ultimate patterns of competing
orders is subtle. For instance, there are striped structures interpreted
in terms of actual or incipient orders~\cite{kivelson16}, or attributed to be induced by lattice distortion or impurity-pinning effect~\cite{normand,yang17}.
To date, the issue of the nature of spatial inhomogeneous phases in cuprates is still
controversial.

Impurity effects have been proven to be a valuable probe to explore
the fundamental properties of cuprates since it often provides important
insights into the underlying physics. Experimentally, STM detection of the
modulation of the local density of states by impurities
was used to great advantage to probe the nature of quasi-particle states of
DSCs both in the superconducting and pseudo-gap phases~\cite{pan,huds21,chatt}.
Another interesting issue is interference patterns and pinning stripes by impurities in DSCs.
In general, impurity substitution causes additional slowing
down of spin fluctuations and pinning of the stripes, leading to
the formation of a static charge or spin order. Zhu et al. have proposed
an explanation to the checkerboard pattern around a impurity or vortex cores based
on an effective mean-field $t-U-V$ model~\cite{zhujx12}. For the
case without an applied magnetic field, the disorder can produce a
similar pinning effect of the fluctuating stripes. Andersen et al.
investigate disorder-induced freezing of incommensurate spin fluctuations,
which agrees qualitatively with experimental
observations~\cite{andersen18}. In the
present work, we study the striped morphologies induced by magnetic impurities
in d-wave superconductors (DSCs) near optimal doping. Including the competitions and
coexistence among the DSC, spin-density wave (SDW), and CDW orders, the system is
explored by self-consistently solving the Bogoliubov - de Gennes
(BdG) equations. It is found that the local disorders can induce
interesting phenomena, including checkerboard, stripe modulations,
and other complex striped morphologies. We expect
that these phenomena could be observed in the STM and NS experiments.

\section{Theory and calculation details}
We start from the two-dimensional $t-t^{\prime}-U-V$ model, which
consists of two parts, $H=H_{0}+H_{imp}$. The Hamiltonian $H_{0}$
and $H_{imp }$ describe the superconductor and magnetic impurities,
respectively, which can be written as
\begin{eqnarray}
H_{0}&=&-\sum_{ij\sigma}t_{ij}(c_{i\sigma}^{\dagger}c_{j
\sigma}+H.c.)  \nonumber \\
& +&\sum_{ij}(\Delta_{ij}c_{i\uparrow}^{\dagger}c_{j\downarrow}^{
\dagger}+H.c.)+\sum_{i\sigma}(Un_{i,\bar{\sigma}}-\mu)c_{i\sigma}^{\dagger}c_{i\sigma},
\nonumber
\\
H_{imp}&=
&\sum_{i}h_{eff}(i)(c_{i\uparrow}^{\dagger}c_{i\uparrow}-c_{i\downarrow}^{
\dagger}c_{i\downarrow}).
\end{eqnarray}
Here $c_{i\sigma}$ annihilates an electron of spin $\sigma$ at the
$i$th site. The hopping integral $t_{ij}$ takes $t$ between nearest
neighbor (NN) pairs, and $t^{\prime}$ between next-nearest neighbor
(NNN) pairs. $U$ is the on-site Coulomb repulsion interaction. $\mu$
is the chemical potential, which is determined self-consistently in
the calculation. The local effective field $h_{eff}(i)$ is
introduced to model the exchange coupling between conducting
electrons and the impurity spin, where we have treated the Kondo
impurity spin as a Ising-like one. Some similar model had been
employed to study the effects of magnetic impurities on cuprate
superconductors~\cite{bala19,zuo20}, which can qualitatively explain
the observed impurity states well~\cite{huds21}. Therefore, we
employed the above model in this work. The self-consistent
mean-field parameters are given by $n_{i}=\sum_{\sigma}
<c_{i\sigma}^{\dagger}c_{i\sigma}>$, the magnetization $m_{i}=(1/2)(<c_{i%
\uparrow}^{\dagger}c_{i\uparrow}>-<c_{i\downarrow}^{\dagger}c_{i\downarrow}>)
$, and the DSC order parameter
$\Delta_{ij}=(V/2)<c_{i\uparrow}c_{j\downarrow}-c_{i\downarrow}c_{j\uparrow}>$
with V the phenomenological pairing interaction.

The Hamiltonian $H$ can be diagonalized by solving the following BdG
equations,
\begin{eqnarray}
\left(
\begin{array}{lr}
H_{ij,\uparrow} & \Delta_{ij} \\
\Delta_{ij}^{\ast} & -H_{ij,\downarrow}^{\ast}
\end{array}
\right)\Psi_{j}=E\Psi_{i},
\end{eqnarray}
where the quasiparticle wave function $\Psi_{i}=(u_{i\uparrow},v_{i
\downarrow})^{T}$. The spin-dependent single-particle Hamiltonian
reads $H_{ij\sigma}=-t\delta_{i+\tau,j}-t^{\prime}\delta_{i+
\tau^{\prime},j}+[\sum_{i_{m}}\sigma
h_{eff}(i)\delta_{i,i_{m}}+Un_{i,\bar{\sigma}}-\mu]\delta_{ij}$.
Here the subscripts $\tau$ and $ \tau^{\prime}$ denote the unit
vector directing along four NN and NNN bonds respectively, and
$i_{m}$ is the position of the impurity site. The self-consistent
parameters are given by $n_{i\uparrow}=\sum_{n}|u^{n}_{i
\uparrow}|^2f(E_{n})$, $n_{i\downarrow}=\sum_{n}|v^{n}_{i
\downarrow}|^2[1-f(E_{n})]$, and $\Delta_{ij}=\frac{V}{4}\sum_{n}[{
u^{n}_{i\uparrow}v_{j\downarrow}^{n\ast}+v_{i\downarrow}^{n\ast}u^{n}_{j
\uparrow}}]tanh(\frac{\beta E_{n}}{2})$, where $f(E)=1/(1+e^{\beta
E})$ is the Fermi-Dirac distribution function. Hereafter, the length
is measured in units of the lattice constant $a$, and the energy in
units of $t$. We set $U=2.5$ and $t^{\prime}=-0.2$ in this paper. The pairing interaction is chosen as $V=1.0$ to
guarantee that the superconducting order $\Delta_{0}\simeq0.08t$,
comparable with the observed $T_{c}$ in cuprate superconductors. We study the impurity effects on the
electronic states of DSCs near optimal doping with the filling
factor $n_{f}=\sum_{i\sigma}c_{i\sigma}^{
\dagger}c_{i\sigma}/(N_{x}N_{y})=0.83$ (i.e., the hole doping
$x=0.17$), where $N_{x}$, $N_{y}$ are the linear dimension of the
unit cell. The BdG equations are solved
self-consistently for a square lattice of $32\times32$ sites, and the periodic boundary conditions are
adopted. The numerical calculation is performed at a very low
temperature, $T=10^{-5}$K, to extract the low-energy physics. The
local effective field is taken to be $h_{eff}$ at the impurity site
and zero otherwise. The DSC order parameter at the $i$th site is
defined as
$\Delta_{i}=(\Delta_{i,i+e_{x}}+\Delta_{i,i-e_{x}}-\Delta_{i,i+e_{y}}-\Delta_{i,i-e_{y}})/4$,
and the spin order parameter is $M_{i}=(-1)^{i}m_{i}$.

\section{Numerical results and discussion}
\begin{figure}[htpb]
  \begin{center}
    \scalebox{1.1}[1.1]{\includegraphics[width=8cm]{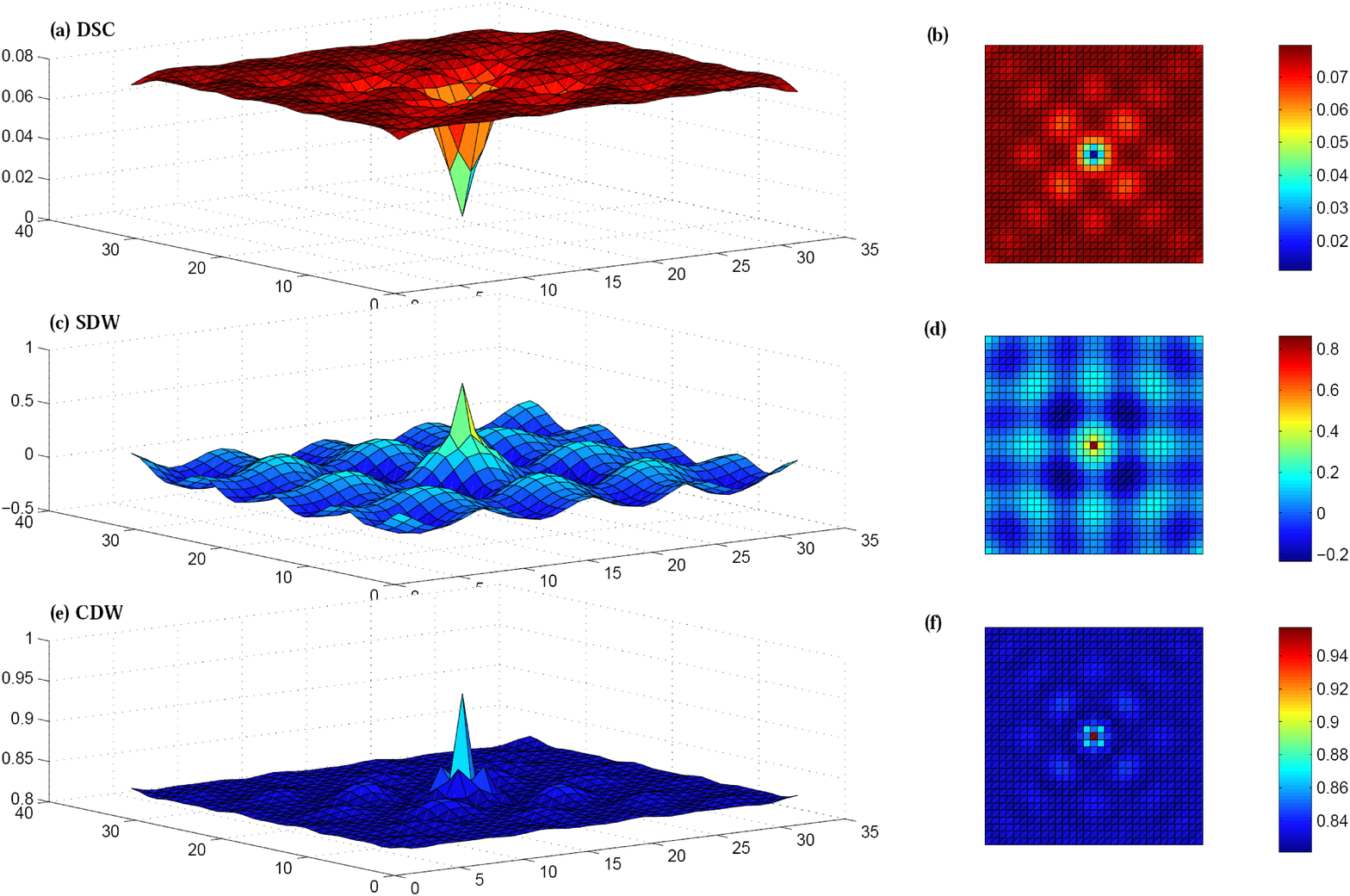}}
  \end{center}
  \caption{(Color online) The surface plots of orders around the
magnetic impurity on a unit cell of size $32\times32$ sites. (a),
(c), and (e) are the spatial distributions of the DSC, spin and
charge orders. (b), (d), and (f) are their contour plots.}
  \label{f.1}
\end{figure}

In FIG.~\ref{f.1}, we plot the spatial distributions of the DSC,
spin, and charge orders around the impurity site. One can see that
all the three orders display checkerboard modulations around the
magnetic impurity. Similar to the nonmagnetic impurity case, a SDW
checkerboard pattern is observed around the
magnetic impurity~\cite{zhujx12,chen22}, which coincides with the NS
data~\cite{lake1}. However, it is noteworthy that the checkerboard
pattern of the DSC order can also be induced by the magnetic
impurity, and a weak associated CDW pattern is observed, which is
different from the nonmagnetic impurity case. Moreover, the
modulated DSC and CDW orders share the same periodicity.
Overall, the DSC order is strongly suppressed at the impurity site
while the amplitudes of the CDW and SDW orders reach global maxima.
This is the common feature of orders around the magnetic impurity
despite various parameters. Therefore, in view of these phenomena,
we clearly see the relationship of competition and coexistence
between antiferromagnetic and DSC orderings. In comparison to the
nonmagnetic impurity case~\cite{zhujx12,chen22}, one finds that in
both cases the DSC orders are suppressed around the impurity site.
However, for the CDW order, the opposite tendency is observed due to
the fact that the nonmagnetic impurity is repulsive.

\begin{figure}[htpb]
  \begin{center}
    \scalebox{1.1}[1.1]{\includegraphics[width=8cm]{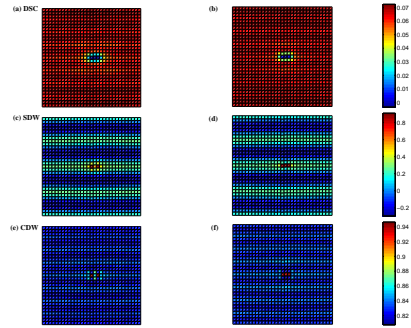}}
  \end{center}
  \caption{(Color online) The spatial contour plots of orders by inserting two or three
magnetic impurities in the same direction. (a),
(c), and (e) are the contour plots of the DSC, spin and
charge orders with two impurities, while (b), (d), and (f) are the three-impurity case.}
  \label{f.2}
\end{figure}

Next we consider the many-impurity case to show the
striped morphologies which could be induced by magnetic impurities in DSCs.
As shown in FIG.~\ref{f.2}, upon
inserting magnetic impurities in the same direction, a local stripe-like structure
can be induced, which is analogous to the nonmagnetic impurity
case~\cite{chen22}. A second important feature is that the average separation between neighboring stripes is not expected to change upon impurity-doping.
The impurity-pinned SDW stripe also show a
modulation with periodicity 8$a$, which coexists with the DSC order.
In addition, similar to the single impurity case, the modulated DSC
and CDW stripes also share the same periodicity. For the three-impurity case,
one notes that actually there are local ferromagnetic moments formed around impurities.
If the three impurities are placed so as to form a right-angled shape, one finds that the striped
structures also show a right-angled shape due to the impurity-pinning and quantum interference effects (see Fig.3).
The impurity-pinned stripes along x or y direction are balanced by quantum interference effects so that
structures with a right-angled shape are favorable on energy. Meanwhile, the alternately dark and bright features due to quantum interference effect
are still obvious. Away from the impurities, the stripes eventually evolve to checkerboard structures due to quasiparticle
interference, and tend to disappear on farther sites. This feature can be seen more clearly from the DSC order parameter [see Fig.3 (a)].

Moreover, we also calculate the cases with four or five impurities. As shown in Fig.~\ref{f.4}, when inserting four or five
magnetic impurities symmetrically around the center site, there are symmetrical impurity-pinned stripe structures formed around impurities.
Interestingly, these stripe structures are somewhat analogous to "quantum corrals", structures built from adatoms using atomic manipulation processes, which have been observed on a metal surface~\cite{crommie23,manoha24}, Here, these quantum-corral-like structures in order parameters are produced due to impurity-pinning and quantum interference effects in DSCs.
This feature is especially clear in the spin order, in which the alternately dark and bright rectangular corrals surround the impurities near the center.
Similar to the three-impurity case, the stripes eventually evolve to checkerboard structures in the sites away from the impurities and disappear on farther sites.

\section{conclusion}
To conclude, we have investigated the modulations
around magnetic impurities of the DSC, spin, and
charge orders in DSCs. It is found that magnetic impurities could induce rich and interesting striped morphologies.
For the single impurity case,
the checkerboard modulations are observed in all three orders.
The weak associated CDW pattern is different from the
nonmagnetic impurity case. When more magnetic impurities are
inserted, more complex modulated structures could be induced, including rectilinear
and right-angled stripes and quantum-corral-like structures due to impurity-pinning and quantum interference effects.
The latter structures were only reported on metal surfaces previously. We expect that these
phenomena could be observed in the high resolution STM and NS
measurements on DSCs.

\section*{Acknowledgements}
This work is supported by the National Nature Science Foundation of China (Grant No. 10904063). Inspiring discussions with Prof. Chang-De Gong and Dr. Yuan Zhou are acknowledged.

\newpage

\begin{figure}[htpb]
  \begin{center}
    \scalebox{1.1}[1.1]{\includegraphics[width=8cm]{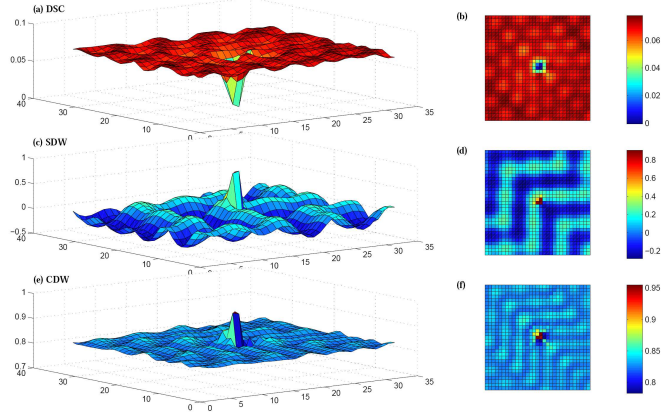}}
  \end{center}
  \caption{(Color online) The same plot as Fig. 1 but with three impurities forming a right-angled shape.}
  \label{f.3}
\end{figure}

\begin{figure}[htpb]
  \begin{center}
    \scalebox{1.1}[1.1]{\includegraphics[width=8cm]{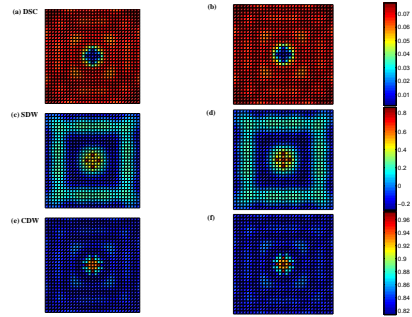}}
  \end{center}
  \caption{(Color online) The spatial contour plots of orders by inserting four or five
magnetic impurities symmetrically around the center site. (a),
(c), and (e) are the contour plots of the DSC, spin and
charge orders with four impurities, while (b), (d), and (f) are the five-impurity case.}
  \label{f.4}
\end{figure}

\end{document}